\RequirePackage[final]{graphicx}
\documentclass[aps, superscriptaddress, prb, twocolumn]{revtex4-2}
\usepackage[utf8]{inputenc}
\usepackage{amsmath}
\usepackage{amssymb}
\usepackage{amsfonts}
\usepackage{mathtools}
\usepackage{fixme}
\usepackage{color}
\usepackage{braket}
\usepackage{dsfont}
\usepackage{hyperref}
\usepackage{url}

\usepackage{tikz}
\usetikzlibrary{calc}

\usetikzlibrary{positioning,arrows}
\usetikzlibrary{decorations}
\usetikzlibrary{decorations.pathmorphing,positioning}
\usetikzlibrary{decorations.markings}

\tikzset
{
hole/.style     = { draw = black, postaction = { decorate }, decoration = { markings, mark = at position .55 with { \arrow[black]{ triangle 45} } } },
spinwave_11/.style    = { draw = cyan, postaction = { decorate }, decoration = { markings, mark = at position .5 with { \arrow[cyan]{ triangle 45} } } },
particle_small_arrow/.style     = { draw = black, postaction = { decorate }, decoration = { markings, mark = at position .5 with { \arrow[scale = 0.6, black]{ triangle 45} } } },
spinwave_12/.style    = { draw = cyan, postaction = { decorate }, decoration = { markings, mark = at position .25 with { \arrow[cyan, >=triangle 45]{<} }, mark = at position .75 with { \arrow[cyan, >=triangle 45]{>} } } },
spinwave_21/.style    = { draw = cyan, postaction = { decorate }, decoration = { markings, mark = at position .25 with { \arrow[cyan, >=triangle 45]{>} }, mark = at position .75 with { \arrow[cyan, >=triangle 45]{<} } } },
spinwave_total/.style       = { decorate, decoration = {snake, amplitude = 2pt, segment length = 7pt } }
}

\tikzset{
  ncbar angle/.initial=90,
  ncbar/.style={
    to path=(\tikqtostart)
    -- ($(\tikqtostart)!#1!\pgfkeysvalueof{/tikz/ncbar angle}:(\tikqtotarget)$)
    -- ($(\tikqtotarget)!($(\tikqtostart)!#1!\pgfkeysvalueof{/tikz/ncbar angle}:(\tikqtotarget)$)!\pgfkeysvalueof{/tikz/ncbar angle}:(\tikqtostart)$)
    -- (\tikqtotarget)
  },
  ncbar/.default=0.5cm,
}

\usepackage{tgtermes} 

\newlength{\reducedsectionspacing}
\newlength{\defaultsectionspacing}
\newlength{\sectiontitlelength}

\newcommand{\shortsection}[2][]{
\section{\setlength{\sectiontitlelength}{\widthof{\thesection\quad}+\widthof{#2}}\ifdim\sectiontitlelength<\linewidth#2
\else
\fontdimen2\font=\reducedsectionspacing
\setlength{\sectiontitlelength}{\widthof{\thesection\quad}+\widthof{#2}}\ifdim\sectiontitlelength<\linewidth#2
\else\fontdimen2\font=\defaultsectionspacing
\ifx\\#1\\#2\else#1
\fi\fi\fi
}}

\tikzset{square left brace/.style={ncbar=0.1cm}}
\tikzset{square right brace/.style={ncbar=-0.1cm}}

\definecolor{myred}{RGB}{214,26,70}
\definecolor{myreddark}{RGB}{76,8,38}
\definecolor{myblue}{RGB}{35,106,185}
\definecolor{mybluedark}{RGB}{19,56,99}
\definecolor{mybluebright}{RGB}{225,236,249}

\def\te{{\rm e}}

\def\bi{{\bf i}}
\def\bj{{\bf j}}

\def\nn{\nonumber}

\def\AF{{ \rm AFM }}

\begin{document}
\title{Omnipresent bound state of two holes in antiferromagnetic Bethe lattices}
\date{\today}

\author{K. \ Knakkergaard \ Nielsen}
\affiliation{Max-Planck Institute for Quantum Optics, Hans-Kopfermann-Str. 1, D-85748 Garching, Germany}

\begin{abstract}
For decades, it has remained an open question, whether two dopants in the $t$--$J$ model bind in the strongly correlated regime of small spin couplings versus hopping, $J \ll t$. Here, we investigate this problem in Bethe lattice structures with Ising coupling $J_z$, and mainly focus on the case of a coordination number equal to 4. The special geometry circumvents subtle effects in regular lattices, but importantly still contains the non-trivial dependency on the rotational symmetry and particle statistics. It, furthermore, allows us to reach numerical convergence, and we conclusively answer whether binding occurs or not. In particular, we find that the rotationally symmetric $s$ waves unbind below $J_z \simeq 0.3 t$, which we unveil is tied to Pauli blocking of symmetrical hole configurations. This is further substantiated by the fact that holes in a bosonic spin environment shows strong -- superlinear -- binding at low $J_z / t$. Additionally, higher rotational fermionic $p$ and $d$ waves partially overcome the blocking, are perfectly degenerate, and bind for all $J_z / t$. In the strongly correlated regime of $J_z \ll t$, this binding occurs sublinearly with scaling $\sim t (J_z / t)^{3/2}$. 
\end{abstract}

\maketitle
\section{Introduction}
The motion of dopants in quantum spin environments is crucial to the understanding of strongly correlated systems \cite{Brinkman1970,Kane1989,Trugman1990}. In particular, the potential binding of two dopants facilitates their pairing instability at finite densities \cite{Randeria1989,Randeria1990}, and should lead to key insights into the microscopic origin of high-temperature superconductivity \cite{highTc}. However, even in effective low-energy models such as the $t$--$J$ model, it remains unsettled whether binding occurs in the limit of strong correlations, $J\ll t$ \cite{Eder1992,Boninsegni1993,Poilblanc1994,Chernyshev1998,Hamer1998,Vidmar2013,Mezzacapo2016,Grusdt2022}. Sparked by new opportunities in quantum simulation experiments with ultracold atoms in optical lattices \cite{Esslinger2010,Bakr2009,Sherson2010,Haller2015,Boll2016,Mazurenko2017,Chiu2018,Yang2021}, this historic problem is now receiving renewed interest \cite{Cheuk2016b,Hilker2017,Brown2019,Chiu2019,Brown2020a,Vijayan2020,Hartke2020,Brown2020b,Gall2021,Koepsell2021}. In particular, the simulation of the Fermi-Hubbard model has revealed spatial correlations in the presence of magnetic polarons \cite{Koepsell2019}, their dynamical formation \cite{Ji2021}, and bound states of two dopants in ladders of one-dimensional chains \cite{Hirthe2022}. In a broader context, polaron quasiparticles appear in a broad range of scenarios where a few impurities interact with a quantum many-body environment \cite{Alexandrov_1994}, and the associated formation of bipolarons generally entails retarded induced interactions, for which detailed or even exact insights are scarce. 

\begin{figure}[t!]
\begin{center}
\includegraphics[width=1.0\columnwidth]{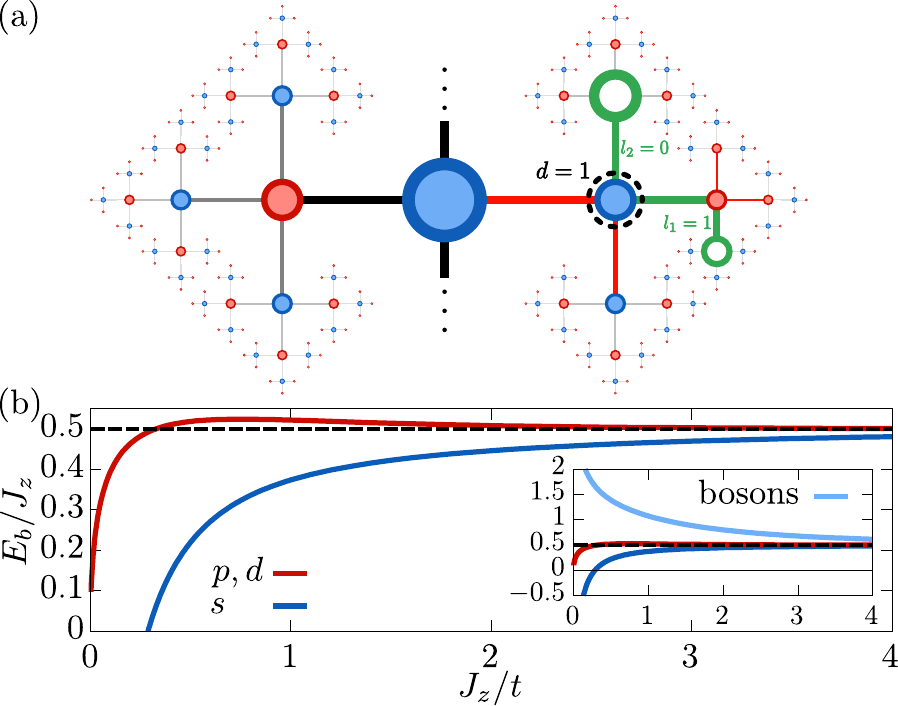}
\end{center}\vspace{-0.5cm}
\caption{(a) Two holes (green) in an antiferromagnetic Bethe lattice of spin--$\uparrow$ (red dots) and spin--$\downarrow$ (blue dots) fermions or hard-core bosons. As the holes separate, a string (green path) of overturned spins between them appears, resulting in frustrated spin bonds (red lines). (b) Two-hole binding energy $E_b = 2E_1 - E_2$ in units of $J_z$ versus $J_z / t$. For weak correlations, $J_z \gg t$, all binding energies approach $J_z / 2$ (dashed line). For $s$ waves, hard-core bosons (light blue in inset) show a superlinear scaling, while fermions (dark blue) unbind below $J_z \simeq 0.3t$. The lowest energy $p$ and $d$ (red) waves are degenerate and always bind.}
\label{fig.binding_energy} 
\vspace{-0.25cm}
\end{figure}

In this Article, we address these issues. We show that in antiferromagnetic Bethe lattices with nearest neighbor hopping $t$ and Ising spin coupling $J_z$, two dopants \emph{always} bind. The underlying mechanism is the emergent, confining string of overturned spins between them \cite{Grusdt2022} [Fig. \ref{fig.binding_energy}(a)]. Interestingly, the binding [Fig. \ref{fig.binding_energy}(b)] highly depends on the particle statistics \cite{Nie2013}. Bosons display a superlinear scaling of the binding energy in the strongly correlated regime of $J_z \ll t$. We reveal that this is facilitated by an enhanced hopping into symmetrical configurations of the holes. For fermions, these configurations are forbidden by the Pauli exclusion principle, highly suppressing the rotationally symmetric $s$-wave states \cite{Chernyshev1999}, and making them unbind below a critical value of $J_z / t$. Higher rotational states may, however, partially overcome the Pauli blocking. Indeed, for four nearest neighbors we find that the lowest $p$ and $d$ waves bind for any value of $J_z / t$, are degenerate, and show a sublinear binding energy for $J_z \ll t$. To obtain these results, we generalize the recent exact treatment of a single hole in Bethe lattices \cite{Nielsen2022_1} to determine the structures possible for the two-hole states at low energies. 

The simplification to the Bethe lattice geometry removes several complications of regular lattices. Most notably, this includes Trugman loops \cite{Trugman1988}, as well as tangential paths which has recently been shown to play a significant role in the incoherent structure of the energy spectrum for a single hole in a two-dimensional Ising antiferromagnet \cite{Wrzosek2021}. However, we emphasize that the unveiled Pauli blocking mechanism is generic, and it may, therefore, still play an important role in regular lattices, even in the presence of isotropic spin couplings. Moreover, it has previously been shown \cite{Chernyshev1999} that the ground state energy for a single hole in an antiferromagnetic Bethe lattice closely follows its regular lattice counterpart. Therefore, in addition to the rare exact insights into the binding of dopants in antiferromagnets, our results may even quantitatively describe the situation in regular lattices with Ising-type spin interactions.

%%%%%%%%%%%%%%%%%%%%%%%%%%%%%%%%%%%%%%%%%%%%%%%%%%%%%%%%%%%%%%%%%%% 
%%%%%%%%%%%%%%%%%%%%%%%%%%%%%%%%%%%%%%%%%%%%%%%%%%%%%%%%%%%%%%%%%%%

The Article is organized as follows. In Sec. \ref{sec.model}, we set up the $t$-$J_z$ model and perform the Holstein-Primakoff transformation to describe the system in terms of holes and spin excitations. In Sec. \ref{sec.rotational_subspaces}, we use the rotational symmetry of the system to split the Hamiltonian into its rotational subspaces, and compute the binding energy in each case. In Sec. \ref{sec.spatial_distribution}, we analyze the relative distribution of the two holes. Finally, we discuss the possible impact of the current results on regular lattice situation in Sec. \ref{sec.discussion}, before we conclude in Sec. \ref{sec.conclusions}.

\section{The model} \label{sec.model}
We consider general antiferromagnetic Bethe lattices of spin-$1/2$ particles, featuring nearest neighbor hopping and Ising-type nearest neighbor spin-spin interactions. The system is described by the $t$--$J_z$ model,
\begin{equation}
\!\!\!\!\hat{H} = \sum_{\braket{\bi, \bj}} \!\!\left[\!-t\sum_\sigma\!\!\left(\tilde{c}^\dagger_{\bi, \sigma}\tilde{c}_{\bj, \sigma} \!+\! {\rm H.c.} \right) \!+\! J_z\!\!\left(\!\hat{S}^{(z)}_\bi\!\hat{S}^{(z)}_\bj \!-\! \frac{\hat{n}_\bi \hat{n}_\bj}{4} \!\right)\!\right]\!,\!
\label{eq.t_Jz_Ham}
\end{equation}
for spin $\sigma = \uparrow,\downarrow$, onsite density operator $\hat{n}_\bj = \sum_\sigma \hat{n}_{\bj, \sigma} = \sum_\sigma\hat{c}^\dagger_{\bj, \sigma}\hat{c}_{\bj, \sigma}$, and with constrained hopping through $\tilde{c}^\dagger_{\bj,\sigma} = \hat{c}^\dagger_{\bj,\sigma}(1 - n_{\bj})$, allowing for up to one spin per site. The isotropic $t$--$J$ model naturally arises as the low-energy description of the Fermi-Hubbard model \cite{Dagotto1994}. The simplification to the $t$--$J_z$ model in Eq. \eqref{eq.t_Jz_Ham} along with the Bethe lattice geometry allows us to determine the fate of the binding. To exactly pinpoint the unbinding mechanism for $s$-wave fermions, we also analyze this situation for hard-core bosons. Here, the spin designates two internal states with the nearest neighbor interaction written in Eq. \eqref{eq.t_Jz_Ham} and pseudospin-$z$ operator $\hat{S}^{(z)}_\bj = (\hat{c}^\dagger_{\bj, \uparrow}\hat{c}_{\bj, \uparrow} - \hat{c}^\dagger_{\bj, \downarrow}\hat{c}_{\bj, \downarrow}) / 2$. We perform a Holstein-Primakoff transformation to obtain a more efficient description in terms of spin excitations and holes, where
\begin{align}
\hat{H}_J = -J_z \sum_{\braket{\bi, \bj}} & \left[\left(\frac{1}{2} - \hat{s}^\dagger_\bi\hat{s}_\bi\right)\left(\frac{1}{2} - \hat{s}^\dagger_\bj\hat{s}_\bj\right) + \frac{1}{4} \right] \nn \\
\times &\left[1-\hat{h}_\bi^\dagger \hat{h}_\bi\right]\left[1-\hat{h}_\bj^\dagger \hat{h}_\bj\right]
\label{eq.H_J_holstein_primakoff}
\end{align}
is the resulting resulting spin coupling, for which the antiferromagnetic ground state satisfies $\hat{s}_\bi\ket{\AF} = 0$. Furthermore, the hopping may be written as 
\begin{align}
\hat{H}_t = t \sum_{\braket{{\bf i}, {\bf j}}} \! \Big[ &\hat{h}^\dagger_{\bf j} F(\hat{h}_{\bf j}, \hat{s}_{\bf j}) F(\hat{h}_{\bf i}, \hat{s}_{\bf i}) \hat{h}_{\bf i} \hat{s}_{\bf j} \nn \\
+ & \hat{s}^\dagger_{\bf i}\hat{h}^\dagger_{\bf j} F(\hat{h}_{\bf j}, \hat{s}_{\bf j}) F(\hat{h}_{\bf i}, \hat{s}_{\bf i}) \hat{h}_{\bf i} \Big] + {\rm H.c.},
\label{eq.H_t_holstein_primakoff}
\end{align}
where $F(\hat{h}, \hat{s}) = \sqrt{1 - \hat{h}^\dagger\hat{h} - \hat{s}^\dagger\hat{s}}$ constrains the motion to ensure at most a single spin excitation on each site. The spin excitations $\hat{s}_\bi$ are bosonic, while the holes $\hat{h}_\bi$ are either fermions or bosons depending on the underlying particle statistics -- fermions and hard-core bosons respectively. Details of the transformation can be found in Appendix \ref{app.HP_transform}.

\section{Rotational subspaces} \label{sec.rotational_subspaces}
For $q$ nearest neighbors, we exploit the discrete rotational $C_q$ symmetry of the system to find simultaneous eigenstates of $\hat{H}$ and the rotation operator $\hat{C}_q$. Rotating $q$ times returns the state back to itself, $(\hat{C}_q)^q \ket{\Psi} = \ket{\Psi}$, giving the eigenvalues $\te^{im 2\pi / q}$, with $m = 0,1,2,\dots,q-1$, and where $m = 0$ corresponds to $s$-wave states. Additionally, it is essential that the low-energy two-hole eigenstates \emph{must} have a nonzero overlap with states with adjacent holes in a perfectly N{\'e}el ordered background. Any other state will at least be elevated by an energy $\propto J_z$. For this reason, even as hopping delocalizes the hole pair, it only disturbs the spin configuration between them [Fig. \ref{fig.binding_energy}(a)]. This allows us to compute the matrix representation of the $t$--$J_z$ Hamiltonian efficiently within each rotational subspace for the low-lying two-hole states, making it possible to go to large system sizes and ensure numerical convergence.

\subsection{$s$ waves} \label{subsec.s_waves}
To understand how Pauli blocking suppresses the bound state for fermions, we first analyze the rotationally symmetric $s$-wave states. To describe the lattice, we pick a site as the origin $0$. Sites $n$ hops away from the origin is said to be at depth $n$ and are denoted $\bj_n = 0,j_1,\dots,j_n$. Here, $j_1 = 1,2,\dots,q$, and $j_l = 1,2,\dots,q-1$ for $l\geq 2$ denote the sites at each depth \cite{Nielsen2022_1,Katsura1974}. Starting from two adjacent holes in a perfect N{\'e}el background, we include all states that can be reached by the hopping Hamiltonian. In this way, two holes may then go onto separate paths of the Bethe lattice at the \emph{divergence depth} $d\geq 0$, and continue to form two strings of overturned spins of length $l_1 \geq 0$ and $l_2 \geq -1$ out to the respective holes [Fig. \ref{fig.binding_energy}(a)]. The corresponding $s$ wave for such a configuration is
\begin{align}
\ket{\Psi^{0}_d(l_1,l_2)} =& \, [N^s_d(l_1,l_2)]^{-1/2}\!\!\!\sum_{\bj_{d+l_1+1}} \!\!\! \sum_{\substack{\bi_{d+l_2+1}:\\\bi_d = \bj_d\\i_{d+1} \neq j_{d+1}}} \!\!\! \hat{h}^\dagger_{\bi_{d + l_2 + 1}} \hat{s}^\dagger_{\bj_{d}} \nn \\
& \times \!\!\!\prod_{k_2=1}^{l_2}\hat{s}^\dagger_{\bi_{d + k_2}} \prod_{k_1=1}^{l_1}\hat{s}^\dagger_{\bj_{d + k_1}} \hat{h}^\dagger_{\bj_{d + l_1 + 1}}\!\ket{\AF},
\label{eq.symmetric_states}
\end{align}
where $N^s_d(l_1,l_2)$ is the total number of possible configurations of the two holes, which defines the normalization [See Appendix \ref{app.s_waves}]. For $l_2 = -1$, the state denotes two holes on the same path in the Bethe lattice. Swapping $l_1,l_2$ leads to an overall sign, $\ket{\Psi^0_d(l_2,l_1)} = \pm \ket{\Psi^0_d(l_1,l_2)}$, with $+$ ($-$) for bosons (fermions). For fermions, this entails that an $s$ wave with the symmetric hole configuration $l_1 = l_2$ cannot be formed, manifesting the Pauli exclusion principle. We, thus, keep $l_2 \!\leq\! l_1$ ($l_2 \!<\! l_1$) for bosons (fermions). For bosons, the symmetric configuration $\ket{\Psi^0_d(l_1,l_1)}$ has an additional factor of $1/\sqrt{2}$ with respect to Eq. \eqref{eq.symmetric_states}.  

The fact that spin excitations only appear between the holes when they depart from each other allows us to define a closed $s$-wave low-energy subspace and calculate the matrix representation of the Hamiltonian [see Appendix \ref{app.s_waves}] to find the lowest energy $s$-wave eigenstate. In particular, the states $\ket{\Psi_d(l_1,l_2)}$ are eigenstates of $\hat{H}_J$ with matrix elements $\bra{\Psi^0_d(l_1,l_2)} \hat{H}_J \ket{\Psi^0_d(l_1,l_2)} = V_J(l_1 + l_2 + 1)$ given by the linear string potential
\begin{align}
V_J(l_s) = l_s\cdot (q-2) \frac{J_z}{2},
\label{eq.string_potential}
\end{align}
for any $l_s \geq 1$. This emergent potential can be understood fairly simply in the following manner. When the holes are separated by a string of $l_s \geq 1$ overturned spins, the spin bonds along the string remains antiferromagnetic and yields no energy cost. However, all spin bonds off the string are now ferromagnetic and, thus, frustrated. Consequently, for $q$ nearest neighbors, there are $q-2$ frustrated spin bonds per overturned spin. Hence, the magnetic energy cost is $q-2$ times the length of the string $l_s$ times the spin-bond energy $J_z / 2$. In the special case of $l_s = 0$, when the holes are adjacent, they share a frustrated spin bond. Since we choose two separate stationary holes as the energy reference -- with energy $qJ_z$ above the antiferromagnetic ground state -- this leads to the value of $V_J(l_s = 0) = - J_z / 2$. 

The hopping Hamiltonian may, furthermore, couple $\ket{\Psi_d^0(l_1,l_2)}$ to its nearest neighbors. In the case when the hopping does not couple to a symmetrical configuration of the holes, we find that the hopping matrix element is increasing by an amount defined by the number of configurations before and after the hop, $\bra{\Psi^0_{d'}(l_1',l_2')} \hat{H}_t \ket{\Psi^0_d(l_1,l_2)} = t \sqrt{N_{d'}^s(l_1',l_2') / N_d^s(l_1,l_2)}$ [See Appendix \ref{app.s_waves}]. For the allowed hopping processes this yields
\begin{align}
\bra{\Psi^0_d(l_1 + 1,l_2)} \hat{H}_t \ket{\Psi^0_d(l_1,l_2)} &= t\sqrt{q-1}, \nn \\
\bra{\Psi^0_{d\geq 1}(l_1,0)} \hat{H}_t \ket{\Psi^0_{d\geq 1}(l_1,-1)} &= t\sqrt{q-2}, \nn \\
\bra{\Psi^0_{d}(l_1,l_2 + 1)} \hat{H}_t \ket{\Psi^0_{d}(l_1,l_2)} &= t\sqrt{q-1}, \nn \\
\bra{\Psi^0_{d-1}(l_1+1,-1)} \hat{H}_t \ket{\Psi^0_{d}(l_1,-1)} &= t.
\label{eq.H_t_matrix_elements_s_wave}
\end{align}
The third line applies only when either $d = 0$ or $l_2 > -1$. For bosons, it is, furthermore, possible to hop out off or into the symmetrical configurations $\ket{\Psi_{d}(l_1,l_1)}$, when $(l_1,l_1) \to (l_1 + 1,l_1)$ and $(l_1,l_1 - 1) \to (l_1,l_1)$. This yields a Bose-enhancement of $\sqrt{2}$ with respect to Eq. \eqref{eq.H_t_matrix_elements_s_wave} for these processes. For $s$-wave fermions, these hopping events are totally absent. 

\begin{figure}[t!]
\begin{center}
\includegraphics[width=1.0\columnwidth]{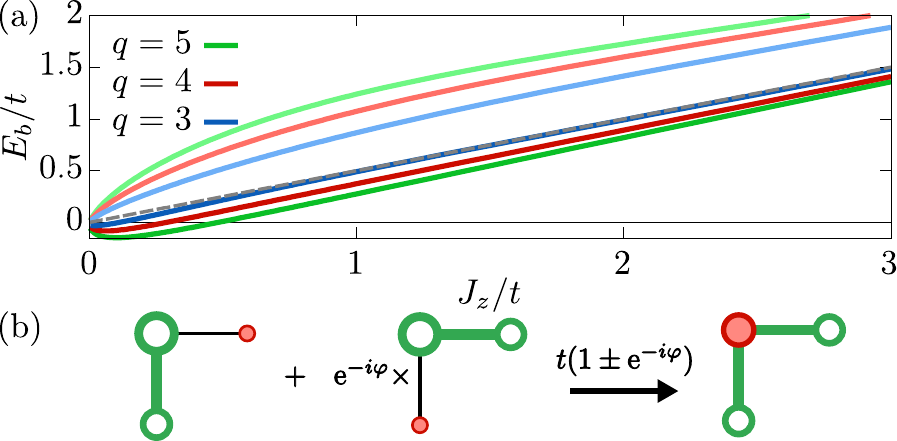}
\end{center}\vspace{-0.5cm}
\caption{(a) Binding energy versus $J_z / t$ for $s$ wave fermions (dark colors) and hard-core bosons (lighter colors) for indicated numbers of nearest neighbors, $q$, and compared to the weakly correlated result $E_b = J_z / 2$ (grey dashed line) approached for $J_z \gg t$. While bosons experience superlinear binding energies at low $J_z / t$, fermions unbind at a critical value of $J_z / t$ increasing with $q$. (b) Origin of unbinding. When the holes hop into a symmetrical configuration, quantum interference takes place between two pathways, depending on their relative phase, $\te^{-i\varphi}$, and their particle statistics ($+$ for bosons, $-$ for fermions), blocking the hopping for $s$ wave fermions ($\varphi = 0$).}
\label{fig.binding_energy_s_wave} 
\vspace{-0.25cm}
\end{figure}

With the matrix representation in place, we find their eigenstates and two-hole energies $E_2$. The binding energy $E_b = 2E_1 - E_2$ [Fig. \ref{fig.binding_energy_s_wave}(a)] is computed by using the exact results for the single-hole ground state energy $E_1$ \cite{Nielsen2022_1}. This explicitly shows that $s$-wave fermions are always less bound than their bosonic counterpart. In particular, the binding energy for $s$-wave fermions are always below the asymptotic $J_z/2$ line, while the bosonic binding energy is above this line. In the strongly correlated regime, $J_z \ll t$, we expect an asymptotic behavior $E_2 \to -4\sqrt{q-1}t + a \cdot t(J_z/t)^{2/3} + b J_z$ in analogy to the single hole case \cite{Nielsen2022_1,Chernyshev1999}. Here, the dominant correction $\propto t(J_z/t)^{2/3}$ stems from the linear potential in Eq. \eqref{eq.string_potential}. A fit to this functional form is performed in Appendix \ref{app.power_law_strong_correlations} and leads to the asymptotic scalings
\begin{align}
E_b^{(s)} &\simeq -1.4 t\left(\frac{J_z}{t}\right)^{\!\!2/3} + 2.4 J_z,\nn\\
E_b^{(b)} &\simeq 0.94 t\left(\frac{J_z}{t}\right)^{\!\!2/3} + 0.39 J_z,
\label{eq.binding_energy_strong_interactions}
\end{align}
for $s$-wave fermions and hard-core bosons, respectively, in the limit $J_z \ll t$. This demonstrates that bosons show superlinear binding energies at low $J_z / t$, while fermions unbind below a critical value. Figure \ref{fig.binding_energy_s_wave}(a) illustrates that this critical value increases with the number of nearest neighbors. We stress that the mechanism behind the unbinding is the \emph{Pauli excluded hopping} into the $l_1 = l_2$ modes, as this is the \emph{only} difference between the two cases. The mechanism is illustrated in Fig. \ref{fig.binding_energy_s_wave}(b) for a relative phase of $\te^{-i\varphi}$ of two states hopping into the same final hole configuration, vanishing for $s$-wave fermions ($\varphi = 0$). This should be contrasted to the frustration effect \cite{Trugman1988}, which hinges on the interchange of spins. These results agree qualitatively with diagrammatic calcuations \cite{Chernyshev1999}.

%%%%%%%%%%%%%%%%%%%%%%%%%%%%%%%%%%%%%%%%%%%%%%%%%%%%%%%%%%%%%%%%%%% 
%%%%%%%%%%%%%%%%%%%%%%%%%%%%%%%%%%%%%%%%%%%%%%%%%%%%%%%%%%%%%%%%%%%
%%%%%%%%%%%%%%%%%%%%%%%%%%%%%%%%%%%%%%%%%%%%%%%%%%%%%%%%%%%%%%%%%%% 
\begin{figure}[t!]
\begin{center}
\includegraphics[width=1.0\columnwidth]{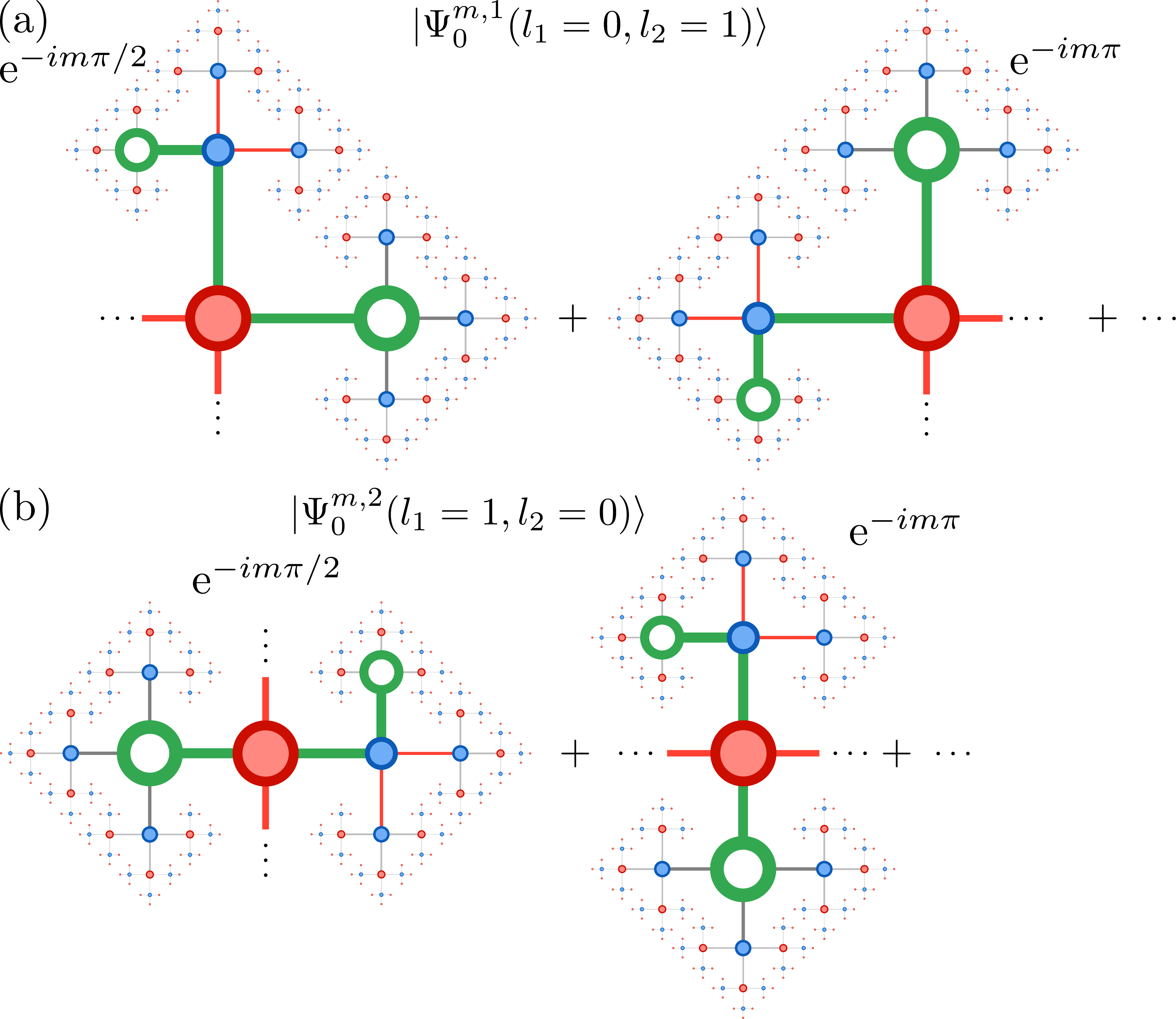}
\end{center}\vspace{-0.5cm}
\caption{Structure of the $p$ ($m = 1$) and $d$ ($m = 2$) waves for holes on neighboring arms (a), and opposite arms (b) of the Bethe lattice. The first two of four terms are shown in each case.}
\label{fig.rotational_states} 
\vspace{-0.25cm}
\end{figure} 
%%%%%%%%%%%%%%%%%%%%%%%%%%%%%%%%%%%%%%%%%%%%%%%%%%%%%%%%%%%%%%%%%%%

\subsection{Higher rotational states} \label{subsec.higher_rotational_states}
We now turn to address, whether higher rotational states can circumvent the Pauli-blocking mechanism and bind for all $J_z / t$. When the holes diverge from each other at $d \geq 1$, or they are on the same path of the lattice ($l_2 = -1$), the states $\ket{\Psi^m_d(l_1,l_2)}$ have the same structure as the $s$-waves in Eq. \eqref{eq.symmetric_states}, only picking up a phase factor ${\rm e}^{-imj_1 \cdot 2\pi/q}$ on each arm $j_1 = 1,\dots,q$ around the origin to secure the $\hat{C}_q$--eigenvalue $\te^{+im \cdot 2\pi / q}$. For this reason, the associated matrix elements are unchanged from the fermionic $s$-wave case, and the Pauli blocking mechanism still takes effect. In fact, almost all matrix elements for the higher rotational states will be identical to the $s$-wave case, including the linear string potential between the holes in Eq. \eqref{eq.string_potential}. However, when the holes diverge at the origin $d = 0$, the relative phases of the arms leads to a partial breaking of the Pauli blocking in Fig. \ref{fig.binding_energy_s_wave}(b). The associated states also turn out to have a much richer structure, because rotationally distinct hole configurations emerge. This even depends sensitively on the number of nearest neighbors. Therefore, we restrict the analysis to $q = 4$, which is geometrically closest to the highly relevant 2D square lattice. Here, $m = 1$ and $m= 2$ give the $p$ and $d$ waves, respectively. For these, we find that the appropriate rotational eigenstates are
\begin{align}
\ket{\Psi^{m,c}_0(l_1,l_2)} = [N_0(l_1,l_2)]^{-1/2}\!\!\sum_{\bj_{l_1+1}}\!\!\!\sum_{\substack{\bi_{l_2+1}:\\i_1 = j_1 + c}} \!\!\!\!\! {\rm e}^{-imj_1 \cdot 2\pi/4}& \nn \\
\times  \hat{h}^\dagger_{\bi_{l_2 + 1}} \hat{s}^\dagger_{0} \!\!\prod_{k_2=1}^{l_2}\hat{s}^\dagger_{\bi_{k_2}} \!\!\prod_{k_1=1}^{l_1}\hat{s}^\dagger_{\bj_{k_1}} \hat{h}^\dagger_{\bj_{l_1 + 1}}\ket{\AF}&,
\label{eq.p_and_d_wave_states_c}
\end{align}
where $c = 1,2$ designate holes on neighboring and opposite arms of the Bethe lattice [see Fig. \ref{fig.rotational_states}]. Also $i_1 = j_1 + c$ is taken modulo $q = 4$ -- i.e. $i_1 = c$ for $j_1 = 4$. For $c = 1$, swapping $l_1\neq l_2$ leads to an orthogonal state, i.e. $\braket{\Psi^{m,1}_0(l_1,l_2)|\Psi^{m,1}_0(l_2,l_1)} = 0$. This means that unlike the $s$-wave case, the $l_2 \geq l_1$ states must be included in the basis. For $c = 2$, on the other hand, the state obtains a sign, $\ket{\Psi^{m,2}_0(l_2,l_1)} = (-1)^{m-1}\ket{\Psi^{m,2}_0(l_1,l_2)}$, depending on whether it is a $p$ wave ($+$) or a $d$ wave ($-$). Therefore, no $d$ wave of the form $\ket{\Psi^{2,2}_0(l_1,l_1)}$ can be formed, while $p$ waves get an additional symmetrization factor of $1/\sqrt{2}$, as for bosons. Furthermore, this (anti)symmetry means that we restrict $l_2 \leq l_1$  ($l_2 < l_1$) for $p$ ($d$) waves. The normalization constant $N_0(l_1,l_2)$ is once again the number of distinct hole configurations and is determined in Appendix \ref{app.p_d_waves}. The relative phases of the arms obtained for these states allow the fermions to overcome the Pauli blocking. In particular, when one of the holes is at the center of the lattice, $l_2 = -1$, this couples to two distinct configurations of holes on neighboring arms
\begin{align}
\bra{\Psi^{m,1}_0(l_1,0)} \hat{H}_t \ket{\Psi^{m}_0(l_1,-1)} &= t,\nn \\
\bra{\Psi^{m,1}_{0}(0,l_1)} \hat{H}_t \ket{\Psi^{m}_{0}(l_1,-1)} &= -t\cdot {\rm e}^{-im\pi/2},
\label{eq.coupling_to_neighboring_arm}
\end{align}
for $l_1 \geq 1$. If the second hole is, furthermore, at a nearest neighbor site, $l_1 = 0$, these two pathways constructively interfere $\bra{\Psi^{m,1}_0(0,0)} \hat{H}_t \ket{\Psi^{m}_0(0,-1)} = t(1 - {\rm e}^{-im\pi/2})$, overcoming the Pauli blocking. This mechanism by itself favors the $d$ waves ($m = 2$), featuring an increased hopping of $t \to 2t$. Finally, matrix elements containing holes on opposite arms of the Bethe lattice, $\ket{\Psi^{m,2}_0(l_1,0)}$, can also beat the Pauli blocking mechanism. This only happens for $p$ waves, for which we obtain
\begin{align}
\bra{\Psi^{1,2}_0(l_1,l_1)} \hat{H}_t \ket{\Psi^{1,2}_0(l_1,l_1-1)} &= \sqrt{2}\cdot \sqrt{3}t,\nn \\
\bra{\Psi^{1,2}_{0}(0,0)} \hat{H}_t \ket{\Psi^{1}_{0}(0,-1)} &= \sqrt{2}t,
\label{eq.coupling_to_opposite_arm}
\end{align}
featuring a Bose-like enhancement factor of $\sqrt{2}$, favoring them over the $s$ and $d$ waves. Here, the upper line applies for $l_1 \geq 1$. From Eqs. \eqref{eq.coupling_to_neighboring_arm} and \eqref{eq.coupling_to_opposite_arm}, it is unclear, what the symmetry of the ground state will be. To settle this, we set up the matrix representation of the $t$--$J_z$ Hamiltonian in the low-energy $p$- and $d$-wave subspaces described by the matrix elements in Eqs. \eqref{eq.string_potential},\eqref{eq.H_t_matrix_elements_s_wave},\eqref{eq.coupling_to_neighboring_arm} and \eqref{eq.coupling_to_opposite_arm}. See Appendix \ref{app.p_d_waves} for the technical details. The results for the binding energy is shown in Fig. \ref{fig.binding_energy}(b), again using the single-hole energy $E_1$ obtained previously \cite{Nielsen2022_1}. Remarkably, we find that the lowest $p$ and $d$ wave are perfectly degenerate and remain bound for all couplings, $J_z/t$. In fact, Fig. \ref{fig.binding_energy}(b) shows that their binding energies are positive and sublinear at low $J_z / t$, whereby the two-hole energy \emph{exactly} matches both the $(J_z/t)^{2/3}$ and the linear $J_z$ term for two single holes. Indeed, a power-law fit in this regime reveals 
\begin{align}
E_{b}^{(p,d)} \simeq 1.55 t \cdot \left(\frac{J_z}{t}\right)^{\!\!3/2},
\label{eq.E_b_p_d_waves}
\end{align}
confirming that the $p$ and $d$ waves always bind. These findings are in qualitative agreement with diagrammatic calculations \cite{Chernyshev1999}, which, however, found a weaker binding energy scaling of $J_z^2 / t$ for low $J_z / t$. It should also be contrasted to the situation in regular lattices, where approximate results for a 2D square lattice \cite{Chernyshev1998,Hamer1998} find the ground state to have $p$-wave symmetry below $J_z \simeq 1.2t$, whereafter $d$-wave symmetry takes over. This strongly suggests that the symmetry of the ground state in regular lattices is a result of quite subtle geometrical effects, such as tangential paths \cite{Wrzosek2021} or loops \cite{Trugman1988}. The latter effect was even considered qualitatively \cite{Chernyshev1998}, and indeed leads to a weak splitting of the $p$ and $d$ waves.

\section{Spatial distribution} \label{sec.spatial_distribution}
\begin{figure}[t!]
\begin{center}
\includegraphics[width=1.0\columnwidth]{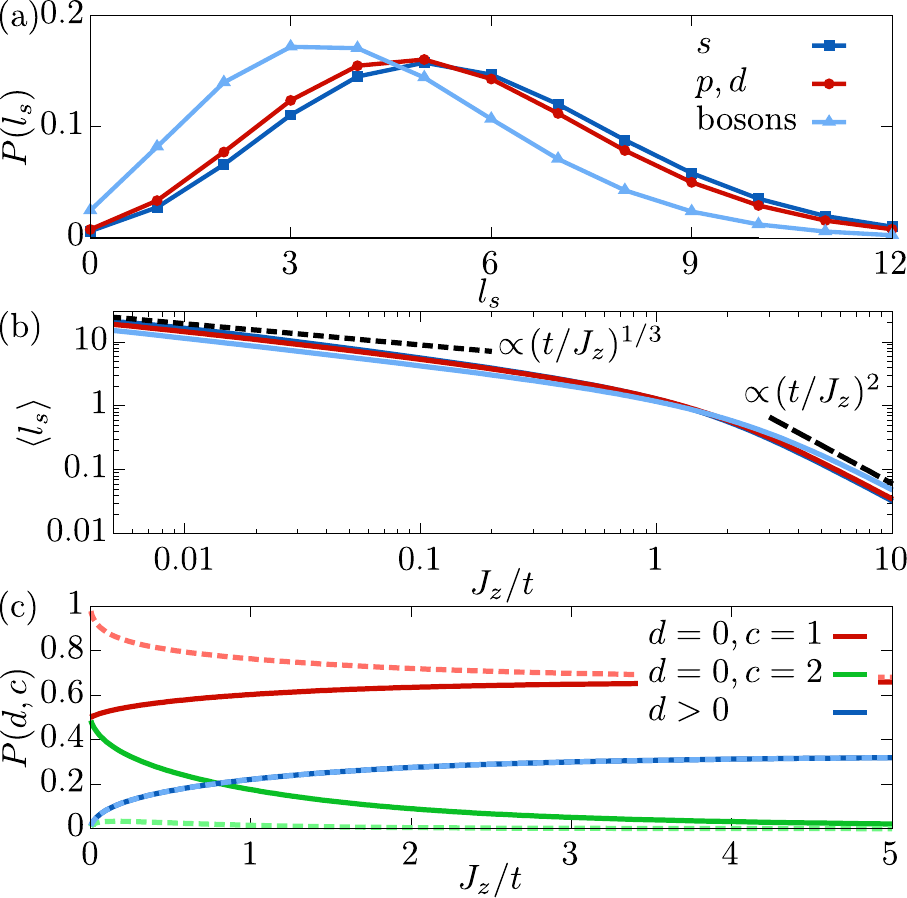}
\end{center}\vspace{-0.5cm}
\caption{(a) String length distribution for $J_z = 0.1t$. (b) Average string length $\braket{l_s} = \sum_{l_s} l_s P(l_s)$ versus $J_z / t$ in a log-log plot. Black lines are guide to the eye for indicated power-law behaviors. (c) Probability to observe the $p$ (full lines) and $d$ (dashed lines) waves in different configurations (colors), with holes on neighboring (opposite) arms of the Bethe lattice for $c = 1$ ($c = 2$) [Fig. \ref{fig.rotational_states}]. The probability for the holes to diverge from each other for $d > 0$ (blue lines) is identical for the two and is always below $40\%$.}
\label{fig.prob_string_length} 
\vspace{-0.25cm}
\end{figure}

To better understand the structure of the two-hole molecular states, we compute the string length probability distribution
\begin{align}
P(l_s) &= \!\!\! \sum_{\substack{d,l_1,l_2: \\ l_1 + l_2 + 1 = l_s}} \!\!\!\!\!\!\!\!\!|\psi_d(l_1,l_2)|^2,
\label{eq.Prob_string_length}
\end{align}
writing the states as $\ket{\Psi} = \sum_{d,l_1,l_2} \psi_d(l_1,l_2)\ket{\Psi_d(l_1,l_2}$. This also corresponds to the distribution of the relative distance $l_s + 1$ of the holes, i.e. the relative wave function, and is plotted in Fig. \ref{fig.prob_string_length}(a) for $J_z = 0.1t$. This exemplifies what we always find to happen: not only are the $p$ and $d$ waves degenerate, their relative wave functions are also identical. To characterize the size of the state versus $J_z / t$, we compute the average string length $\braket{l_s} = \sum_{l_s}l_s P(l_s)$ in Fig. \ref{fig.prob_string_length}(b). We find that the fermionic $s$ wave differs only slightly from the $p$ and $d$ waves, whereas the bosonic $s$ wave is the largest for $J_z \gg t$, and the smallest for $J_z \ll t$. In the former regime, perturbation theory dictates that the string length scales as $(t/J_z)^2$. Conversely, for $J_z \ll t$, the linear string potential in Eq. \eqref{eq.string_potential} leads to an effective length scale of order $(t/J_z)^{1/3}$ \cite{Nielsen2022_1}. Although the relative wave functions for $p$ and $d$ waves are identical, their spatial shapes are markedly different as shown in Fig. \ref{fig.prob_string_length}(c). While the holes generally prefer to be on neighboring arms of the Bethe lattice, the probability to observe the holes on \emph{opposite arms} in the $p$-wave case is enhanced towards the strongly correlated regime $J_z \ll t$, whereas the probability to find the $d$-wave holes on \emph{neighboring arms} increases towards unity. This can be understood from Eqs. \eqref{eq.coupling_to_neighboring_arm} and \eqref{eq.coupling_to_opposite_arm}, which dictate that the $p$ wave experiences enhanced hopping into symmetrical hole configurations in both configurations, whereas this only happens for the $d$ wave, when the holes are on neighboring arms of the Bethe lattice. It is quite remarkable that two states that distribute the holes in such different ways lead to the same relative wave function and degenerate energies. 

%%%%%%%%%%%%%%%%%%%%%%%%%%%%%%%%%%%%%%%%%%%%%%%%%%%%%%%%%%%%%%%%%%% 
%%%%%%%%%%%%%%%%%%%%%%%%%%%%%%%%%%%%%%%%%%%%%%%%%%%%%%%%%%%%%%%%%%%

\section{Discussion} \label{sec.discussion}
In this section, we discuss what bearing the current results have on our understanding of the situation in regular lattices. On one hand, previous results for Ising spin interactions \cite{Chernyshev1999} have shown a very good correspondence between the ground state energy for a single hole in a Bethe lattice with coordination number $q = 4$ and the two-dimensional square lattice over a wide parameter range. This connection really only breaks down at extremely low values of $J_z / t$, where the Nagaoka effect becomes significant \cite{Nagaoka1966} leading to an increasingly spin-polarized background \cite{White2001}. By extension, it could very well be the case that the ground state for two holes inherit this nice correspondence. However, on the other hand, the binding energy found in Eq. \eqref{eq.E_b_p_d_waves} comes at a non-trivially high order in $J_z/t$, suggesting that the holes only marginally bind even in the currently investigated Bethe lattice structure. Since the additional geometric effects \cite{Wrzosek2021,Trugman1988} make the separate holes more mobile in regular lattices, we can expect them to bind more weakly here. Perhaps even more importantly, the presence of flip-flop spin interactions in the isotropic Heisenberg model leads to breaking of the confining string between the holes. Therefore, any binding facilitated by this geometric string in regular lattices, if present \cite{Eder1992,Boninsegni1993,Poilblanc1994,Chernyshev1998,Hamer1998,Vidmar2013,Mezzacapo2016,Grusdt2022}, should be very weak. Finally, recent investigations for isotropic spin couplings \cite{Zhao2022} suggest an alternative binding mechanism via the phase string effect \cite{Sheng1996}, rendering the phenomenolgy of the binding mechanism even more unclear in this case. Therefore, it is uncertain whether the results achieved in these idealized Bethe lattice structures can qualitatively describe the situation in the isotropic $t$--$J$ model in regular lattices. Once again, however, we emphasize that the unveiled Pauli blocking effect is not sensitive to the binding mechanism, and may, therefore, still play a role. 

\section{Conclusions} \label{sec.conclusions}
We analyzed the low-energy two-hole eigenstates in antiferromagnetic Bethe lattices for both bosons and fermions, focusing mostly on the case of four nearest neighbors. This conclusively settles whether a bound state is supported for any $J_z / t$. In particular, we found that the binding dramatically depends on the statistics of the particles. While bosons feature a superlinear binding energy at low values of $J_z / t$, $s$-wave fermions become unbound. We tracked this back to Pauli blocking of hopping into symmetrical hole configurations. In turn, this facilitates that the higher rotational $p$ and $d$ waves become favorable, as they partially overcome the Pauli blocking. We found them to be perfectly degenerate and show a \emph{sublinear} binding energy $\sim t(J_z / t)^{3/2}$ in the strongly correlated regime, $J_z/t\ll 1$. They also have identical relative wave functions, but differ in how the holes distribute over the arms of the Bethe lattice. 

While the Bethe lattice structure is crucially important to derive our numerically exact results, we emphasize that the underlying mechanisms are generic. Indeed, the revealed Pauli blocking mechanism takes place no matter the shape of the lattice, and should play a role in regular lattices as well. Moreover, previous studies \cite{Chernyshev1999} have shown that the Bethe lattice description for the ground state of a single hole closely follows the exact diagonalization results for a hole in a 2D square lattice. This happens even though the spectrum of higher-lying states are markedly different in the two cases, due to self-tangential paths and loops in regular lattices \cite{Wrzosek2021,Trugman1988}. In addition to the phenomenological importance of our work and the exact insights into the binding of holes in antiferromagnets, we, therefore, believe that the results may also be descriptive of the regular lattice situation in the case of Ising spin interactions. Finally, we note that the Ising-type spin interactions investigated here, may be realized using polar molecules \cite{Gorshkov2011_2} and Rydberg-dressed atoms in optical lattices \cite{Glaetzle2015,Bijnen2015,Zeiher2016,Zeiher2017,Borish2020,Sanchez2021}. In the future, we hope to explore similar geometries more accessible to such experiments, in which very efficient or even exact descriptions can be carried out to gain further insights into the intriguing and elusive mechanisms behind high-$T_c$ superconductivity. 

\begin{acknowledgements}
The author thanks Marton Kanasz-Nagy, J. Ignacio Cirac, Georg M. Bruun, Jens Havgaard Nyhegn, Timon Hilker, and Pavel Kos for valuable discussions, and Alexander Chernyshev for important input on the manuscript. This Article was supported by the Carlsberg Foundation through a Carlsberg Internationalisation Fellowship. 
\end{acknowledgements}

\appendix
\onecolumngrid

\section{Holstein-Primakoff transformation} \label{app.HP_transform}
In this section, we give a brief description of the Holstein-Primakoff transformation. We start from the $t$--$J_z$ Hamiltonian in Eq. \eqref{eq.t_Jz_Ham},
\begin{equation}
\hat{H} = \sum_{\braket{\bi, \bj}}\left[-t\sum_\sigma\left(\tilde{c}^\dagger_{\bi, \sigma}\tilde{c}_{\bj, \sigma} + {\rm H.c.} \right) + J_z\left(\hat{S}^{(z)}_\bi \hat{S}^{(z)}_\bj - \frac{\hat{n}_\bi \hat{n}_\bj}{4} \right)\right],
\label{app_eq.t_Jz_Ham}
\end{equation}
where $\hat{S}^{(z)}_\bj = (\hat{c}^\dagger_{\bj, \uparrow}\hat{c}_{\bj, \uparrow} - \hat{c}^\dagger_{\bj, \downarrow}\hat{c}_{\bj, \downarrow}) / 2$ is the spin-$z$ operator, and with constrained hopping through $\tilde{c}^\dagger_{\bj,\sigma} = \hat{c}^\dagger_{\bj,\sigma}(1 - n_{\bj})$. The Holstein-Primakoff transformation then amounts to letting
\begin{align}
{\rm {\bf A}}: \hat{S}^{-}_{\bi} &= \hat{s}_{\bi}^\dagger F(\hat{h}_{\bi}, \hat{s}_{\bi}), \; \tilde{c}_{{\bi},\downarrow}=\hat{h}^\dagger_{\bi}\hat{S}_{\bi}^{+}, \nn \\
\tilde{c}_{{\bi},\uparrow} &=\hat h^\dagger_{\bi} F(\hat{h}_{\bi}, \hat{s}_{\bi}), \; \hat{S}_{\bi}^z= \left[\frac{1}{2} - \hat{s}^\dagger_{\bi}\hat{s}_{\bi}\right] [1-\hat{h}_{\bi}^\dagger \hat{h}_{\bi}]. \nn \\
{\rm {\bf B}}: \hat{S}^{+}_{\bj} &= \hat{s}_{\bj}^\dagger F(\hat{h}_{\bj}, \hat{s}_{\bj}), \; \tilde{c}_{{\bj},\uparrow}=\hat{h}^\dagger_{\bj}\hat{S}_{\bj}^{-}, \nn \\
\tilde{c}_{{\bj},\downarrow} &= \hat h^\dagger_{\bj} F(\hat{h}_{\bj}, \hat{s}_{\bj}), \; \hat{S}_{\bj}^z= \left[\hat{s}^\dagger_{\bj}\hat{s}_{\bj} - \frac{1}{2}\right] [1-\hat{h}_{\bj}^\dagger \hat{h}_{\bj}],
\label{app_eq.HP_transform}
\end{align}
on sublattices A and B, respectively. Here, $\hat{s}_{\bi}^\dagger$ ($\hat h^\dagger_{\bi}$) creates a spin excitation (hole) on site $\bi$, and $F(\hat{s},\hat{h}) = \sqrt{1 - \hat{s}^\dagger \hat{s} - \hat{h}^\dagger \hat{h}}$ makes sure that when applying these operators, we remain in the physical Hilbert space with the correct matrix elements. The spin excitations are always bosonic, while the holes are bosons if the spins are hard-core bosons, and fermions if the spins are fermions. One can then check that the algebra generated by the transformed operators in Eq. \eqref{app_eq.HP_transform} is the same as the original operators within the physical Hilbert space, i.e. the space that locally has the three possible configurations $\{\ket{\AF}, \hat{s}^\dagger_\bi\ket{\AF}, \hat{h}^\dagger_\bi\ket{\AF}\}$. The spin coupling is then straightforwardly obtained to be
\begin{align}
\hat{H}_J &= J_z \sum_{\braket{\bi, \bj}}\left[\hat{S}^{(z)}_\bi \hat{S}^{(z)}_\bj - \frac{\hat{n}_\bi \hat{n}_\bj}{4} \right] = -J_z \sum_{\braket{\bi, \bj}} \left[\left(\frac{1}{2} - \hat{s}^\dagger_\bi\hat{s}_\bi\right)\left(\frac{1}{2} - \hat{s}^\dagger_\bj\hat{s}_\bj\right) + \frac{1}{4} \right] \left[1-\hat{h}_\bi^\dagger \hat{h}_\bi\right]\left[1-\hat{h}_\bj^\dagger \hat{h}_\bj\right],
\label{app_eq.Ham_Jz}
\end{align}
using that the density operator is simply $\hat n_\bi = 1 - \hat h_\bi^\dagger \hat h_\bi$. The hopping Hamiltonian is a bit more lenghthy. It reads
\begin{align}
\hat{H} &= -t\sum_{\braket{\bi, \bj}}\left[\tilde{c}^\dagger_{\bi, \uparrow}\tilde{c}_{\bj, \uparrow} + \tilde{c}^\dagger_{\bi, \downarrow}\tilde{c}_{\bj, \downarrow} + {\rm H.c.} \right] = -t\sum_{\braket{\bi, \bj}}\left[F(\hat{h}_{\bi}, \hat{s}_{\bi}) \hat h_{\bi} \hat h^\dagger_{\bj} F(\hat{h}_{\bj}, \hat{s}_{\bj}) \hat s_\bj + \hat{s}_{\bi}^\dagger F(\hat{h}_{\bi}, \hat{s}_{\bi}) \hat{h}_\bi \hat h^\dagger_{\bj} F(\hat{h}_{\bj}, \hat{s}_{\bj}) + {\rm H.c.} \right] \nn \\
&= \mp t\sum_{\braket{\bi, \bj}}\left[\hat h^\dagger_{\bj} F(\hat{h}_{\bi}, \hat{s}_{\bi}) F(\hat{h}_{\bj}, \hat{s}_{\bj}) \hat h_{\bi} \hat s_\bj + \hat{s}_{\bi}^\dagger \hat h^\dagger_{\bj} F(\hat{h}_{\bi}, \hat{s}_{\bi}) F(\hat{h}_{\bj}, \hat{s}_{\bj}) \hat{h}_\bi  + {\rm H.c.} \right] \nn \\
&\to t\sum_{\braket{\bi, \bj}}\left[\hat h^\dagger_{\bj} F(\hat{h}_{\bi}, \hat{s}_{\bi}) F(\hat{h}_{\bj}, \hat{s}_{\bj}) \hat h_{\bi} \hat s_\bj + \hat{s}_{\bi}^\dagger \hat h^\dagger_{\bj} F(\hat{h}_{\bi}, \hat{s}_{\bi}) F(\hat{h}_{\bj}, \hat{s}_{\bj}) \hat{h}_\bi  + {\rm H.c.} \right].
\label{app_eq.Ham_t_}
\end{align}
Here, the overall sign in the second line depends on whether the holes are bosons ($-$) or fermions ($+$). To eliminate this difference, we perform a local gauge transformation $\hat{h}_\bi \to - \hat{h}_\bi$ in the bosonic case. The bottom line above is, hereby, identical to Eq. \eqref{eq.H_t_holstein_primakoff}. With this, we have performed the Holstein-Primakoff transformation both for fermions and bosons. 

\section{$s$ waves} \label{app.s_waves}
In this section, we set up the $s$-wave states, define ordered bases for fermions and bosons, and compute the associated matrix representation of the Hamiltonian. We start from the definition of the $s$-wave states in Eq. \eqref{eq.symmetric_states}
\begin{align}
\ket{\Psi^0_d(l_1,l_2)} =& \, [N^s_d(l_1,l_2)]^{-1/2}\!\!\!\sum_{\bj_{d+l_1+1}} \!\!\! \sum_{\substack{\bi_{d+l_2+1}:\\\bi_d = \bj_d\\i_{d+1} \neq j_{d+1}}} \!\!\! \hat{h}^\dagger_{\bi_{d + l_2 + 1}} \hat{s}^\dagger_{\bj_{d}}\prod_{k_2=1}^{l_2}\hat{s}^\dagger_{\bi_{d + k_2}} \prod_{k_1=1}^{l_1}\hat{s}^\dagger_{\bj_{d + k_1}} \hat{h}^\dagger_{\bj_{d + l_1 + 1}}\ket{\AF},
\label{app_eq.symmetric_states}
\end{align}
describing two holes diverging at the depth $d$ and generating a string of length $l_1 + l_2 + 1$ between them. We note that in the special case of $l_2 = -1$, the states take on a slightly simpler form
\begin{align}
\ket{\Psi^0_d(l_1,-1)} =& \, [N^s_d(l_1,-1)]^{-1/2}\!\!\!\sum_{\bj_{d+l_1+1}} \!\!\! \hat{h}^\dagger_{\bi_{d}} \prod_{k_1=1}^{l_1}\hat{s}^\dagger_{\bj_{d + k_1}} \hat{h}^\dagger_{\bj_{d + l_1 + 1}}\ket{\AF},
\label{app_eq.symmetric_states_l2_minus_1}
\end{align}
For $l_2 \geq 0$, normalization yields
\begin{align}
1 &= \braket{\Psi^0_d(l_1,l_2)|\Psi^0_d(l_1,l_2)} = [N^s_d(l_1,l_2)]^{-1}\!\!\!\sum_{\bj_{d+l_1+1}} \!\!\! \sum_{\substack{\bi_{d+l_2+1}:\\\bi_d = \bj_d\\i_{d+1} \neq j_{d+1}}} \!\!\! 1 \Rightarrow N^s_d(l_1,l_2) = \sum_{\bj_{d+l_1+1}} \!\!\! \sum_{\substack{\bi_{d+l_2+1}:\\\bi_d = \bj_d\\i_{d+1} \neq j_{d+1}}} \!\!\! 1 \Rightarrow \nn \\
N^s_d(l_1,l_2) &= q(q-1)^{d + l_1}\left[\delta_{d,0} (q - 1)^{l_2 + 1} + (1 - \delta_{d,0}) (q - 2) (q - 1)^{l_2}\right].
\label{app_eq.N_d_l1_l2_1}
\end{align}
For $l_2 = -1$, we similarly get $N^s_d(l_1,-1) = \sum_{\bj_{d+l_1+1}} 1 = q(q-1)^{d + l_1}$. They can, thus, be combined to write the $s$-wave normalization constant
\begin{align}
N^s_d(l_1,l_2) &= q(q-1)^{d + l_1}\left[\left(\delta_{d,0} + \delta_{l_2,-1}(1 - \delta_{d,0})\right) (q - 1)^{l_2 + 1} + (1 - \delta_{d,0})(1 - \delta_{l_2,-1}) (q - 2) (q - 1)^{l_2}\right].
\label{app_eq.N_d_l1_l2_2}
\end{align}
This calculation simply amounts to counting the number of possible positions $\bj_{d+l_1+1}$ of one hole, and then the number of positions $\bi_{d+l_2+1}$ given the constraints that $\bi_d = \bj_d$ and $i_{d+1} \neq j_{d+1}$. Furthermore, for bosons, there is the possibility of the symmetric configuration $l_1 = l_2$. Here, the expression in Eq. \eqref{app_eq.symmetric_states} double counts the possible configurations. Keeping the same formula for $N^s_d(l_1,l_2)$ as above in the case of $l_2 = l_1$, this means that an additional factor of $1/\sqrt{2}$ is needed for $\ket{\Psi^0_d(l_1,l_1)}$. As written in the main text, we must restrict $l_1 \leq l_2$ for bosons and $l_1 < l_2$ for fermions. We are now ready to set up the ordered bases
\begin{align}
{\bf Fermions}\!: \big\{&\ket{\Psi^0_0(0,-1)},\ket{\Psi^0_1(0,-1)},\ket{\Psi^0_0(1,-1)}, \ket{\Psi^0_0(1, 0)}, \ket{\Psi^0_2(0,-1)}, \nn \\ 
&\ket{\Psi^0_1(1,-1)}, \ket{\Psi^0_1(1,0)}, \ket{\Psi^0_0(2,-1)}, \ket{\Psi^0_0(2,0)}, \ket{\Psi^0_0(2,1)}, .. \big\}. \nn \\
{\bf Bosons}\!: \big\{&\ket{\Psi^0_0(0,-1)}, \ket{\Psi^0_0(0,0)}, \ket{\Psi^0_1(0,-1)}, \ket{\Psi^0_1(0,0)}, \ket{\Psi^0_0(1,-1)}, \ket{\Psi^0_0(1, 0)}, \ket{\Psi^0_0(1, 1)}, \ket{\Psi^0_2(0,-1)}, \ket{\Psi^0_2(0,0)},  \nn \\ 
&\ket{\Psi^0_1(1,-1)}, \ket{\Psi^0_1(1,0)}, \ket{\Psi^0_1(1, 1)}, \ket{\Psi^0_0(2,-1)}, \ket{\Psi^0_0(2,0)}, \ket{\Psi^0_0(2,1)}, \ket{\Psi^0_0(2,2)}, .. \big\}.
\label{app_eq.ordered_bases}
\end{align}
Here, we make overall blocks that have a constant value $d + l_1$ and running values of $l_1,l_2$ and subblocks with constant $l_1$ and running $l_2$. Note that the basis for the bosonic case is larger, since $l_2 = l_1$ is allowed. The basis is truncated by allowing a maximum value of $n = \max(d + l_1)$, setting a finite size of the system. \\

Importantly, to efficiently set up the matrix representation, we also need an entry formula that relates the triplet $(d,l_1,l_2)$ to the entry in the ordered basis above. For this reason, we need to count how many values of $l_2$ that are for each $l_1$ and how many values of $l_1$ for each $d + l_1$. This yields the entry formulae
\begin{align}
{\bf Fermions}\!: \; &j_s(d,l_1,l_2) = \frac{(d + l_1)(d + l_1 + 1)(d + l_1 + 2)}{6} + \frac{l_1(l_1 + 1)}{2} + l_2 + 2. \nn \\
{\bf Bosons}\!: \; &j_b(d,l_1,l_2) = \frac{(d + l_1)(d + l_1 + 1)(d + l_1 + 2)}{6} + \frac{(d + l_1)(d + l_1 + 1)}{2} + \frac{l_1(l_1 + 3)}{2} + l_2 + 2.
\label{app_eq.entry_formula_s_wave}
\end{align}
We, hereby, define the Hamiltonian matrix as
\begin{equation}
\mathcal{H}[j(d,l_1,l_2), j(d',l_1',l_2')] =  \bra{\Psi^0_{d}(l_1, l_2)} \hat{H}\ket{\Psi^0_{d'}(l_1', l_2')}.
\label{app_eq.matrix_representation}
\end{equation}
The total size of the matrix is $N\times N$, with $N = j_s(0,n,n-1)$ for fermions and $N = j_b(0,n,n)$ for bosons. We usually work with $n = 32$, yielding $N = 6545$ ($N = 7106$) in the fermionic (bosonic) case. \\ 

We now compute the matrix elements for the $s$ waves. In total, we find that a specific state $\ket{\Psi^0_d(l_1,l_2)}$ has nonzero matrix elements with
\begin{align}
{\bf l_2 \geq 0:}\, &\ket{\Psi^0_d(l_1,l_2)},\, \ket{\Psi^0_d(l_1\pm 1,l_2)}, \,\ket{\Psi^0_d(l_1,l_2\pm 1)},\nn \\
{\bf l_2 = -1:}\, &\ket{\Psi^0_d(l_1,-1)},\, \ket{\Psi^0_d(l_1\pm 1,-1)}, \,\ket{\Psi^0_{d\pm 1}(l_1\mp 1,-1)}.
\end{align}
This means that the number of nonzero elements in the Hamiltonian matrix of size $N\times N$ is less than $6N$, defining a sparse matrix that can be diagonalized very efficiently. The diagonal coupling to $\ket{\Psi^0_d(l_1,l_2)}$ comes about, because $\ket{\Psi^0_d(l_1,l_2)}$ is an eigenstate of the spin-spin coupling $\hat{H}_J$. The matrix element is, thus,
\begin{align}
\bra{\Psi^0_d(l_1,l_2)} \hat{H}_J \ket{\Psi^0_d(l_1,l_2)} = V_J(l_1 + l_2 + 1),
\end{align}
with $V_J(l_s) = l_s (q - 2) J_z / 2$ for $l_s \geq 1$ and $V_J(l_s = 0) = - J_z / 2$. When the holes are adjacent, we have $l_2 = -1, l_1 = 0$, and $V_J(l_1 + l_2 + 1) = - J_z / 2$ because the holes share one frustrated spin bond. Here, we should remember that the reference energy is that of two separate stationary holes. When the holes are separated by a string of $l_s$ overturned spins, the spin bonds along the string are still satisfied, but are frustrated off the string. For $q$ nearest neighbors, there are, thus, $q - 2$ frustrated spin bonds per overturned spin. Hence, the magnetic energy cost is $q - 2$ times the length of the string $l_s$ times the spin-bond energy $J_z / 2$. \\

Next, we turn to the states coupled by the nearest neighbor hopping Hamiltonian $\hat{H}_t$. Naturally, this allows only for coupling with states, where one of the holes is \emph{once} removed from the position in $\ket{\Psi^0_d(l_1,l_2)}$. Hence, there are nonzero matrix elements only with the states $\ket{\Psi^0_d(l_1 \pm 1,l_2)}$, $\ket{\Psi^0_d(l_1,l_2\pm 1)}$, and in the particular case of $l_2 = -1$: $\ket{\Psi^0_{d\mp 1}(l_1\pm 1,-1)}$. The explicit matrix elements are now found. First, for a singular change in $l_1 > l_2$
\begin{align}
\!\!\!\bra{\Psi^0_d(l_1 + 1,l_2)} \hat{H}_t \ket{\Psi^0_d(l_1,l_2)} = t \sqrt{\frac{N^s_d(l_1 + 1,l_2)}{N^s_d(l_1,l_2)}}\braket{\Psi^0_d(l_1 + 1,l_2)|\Psi^0_d(l_1 + 1,l_2)} = t \sqrt{\frac{N^s_d(l_1 + 1,l_2)}{N^s_d(l_1,l_2)}}  = t \sqrt{q - 1} ,
\label{app_eq.l_1_change_matrix_element_s_wave}
\end{align}
giving the square root of the ratio of number of configurations $N^s_d(l_1,l_2)$ times the underlying hopping amplitude $t$. From Eq. \eqref{app_eq.N_d_l1_l2_2}, it follows that the ratio is always $q - 1$. Furthermore, in the bosonic case and for $l_1 = l_2$, a similar calculation yields
\begin{align}
\bra{\Psi^0_d(l_1 + 1,l_1)} \hat{H}_t \ket{\Psi^0_d(l_1,l_1)} = 2t \cdot \sqrt{\frac{N^s_d(l_1 + 1,l_1)}{2 N^s_d(l_1,l_1)}} \braket{\Psi^0_d(l_1 + 1,l_1)|\Psi^0_d(l_1 + 1,l_1)} = \sqrt{2} \cdot \sqrt{q - 1} t.
\label{app_eq.l_1_change_matrix_element_s_wave_bosonic_enhancement}
\end{align}
The additional factor of $2$ comes from the fact that starting from $\ket{\Psi^0_d(l_1,l_1)}$, the motion of either hole deeper into the lattice leads to the state $\ket{\Psi^0_d(l_1 + 1,l_1)}$. A Bose-enhancement factor of $\sqrt{2}$ is, hereby, obtained. Similarly, for $l_2 < l_1 - 1$
\begin{align}
\bra{\Psi^0_d(l_1,l_2 + 1)} \hat{H}_t \ket{\Psi^0_d(l_1,l_2)} = t \sqrt{\frac{N^s_d(l_1,l_2 + 1)}{N^s_d(l_1,l_2)}}  = \left\{\begin{matrix} t\sqrt{q-2}, & d \geq 1 \text{ and } l_2 = -1, \\t\sqrt{q-1}, & \text{otherwise}.\end{matrix}\right.
\label{app_eq.l_2_change_matrix_element_s_wave}
\end{align}
The cases arise from using Eq. \eqref{app_eq.N_d_l1_l2_2}. In the case of $l_2 = l_1 - 1$, this hopping matrix element is also enhanced by a factor of $\sqrt{2}$ for bosons. Finally, when $l_2 = -1$, the shallower hole can also hop along the path of the deeper hole. I.e., 
\begin{align}
\bra{\Psi^0_{d - 1}(l_1 + 1, -1)} \hat{H}_t \ket{\Psi^0_d(l_1,-1)} = t \sqrt{\frac{N^s_{d-1}(l_1 + 1,-1)}{N^s_d(l_1,-1)}} = t.
\label{app_eq.d_and_l1_change_matrix_element_s_wave}
\end{align}
This simply yields $t$, because neither $l_2$, nor $d + l_1$ changes. Collectively, Eqs. \eqref{app_eq.l_1_change_matrix_element_s_wave}, \eqref{app_eq.l_1_change_matrix_element_s_wave_bosonic_enhancement}, \eqref{app_eq.l_2_change_matrix_element_s_wave}, and \eqref{app_eq.d_and_l1_change_matrix_element_s_wave} along with their adjoints give all possible hopping matrix elements. Below, we show the first $5\times 5$ entries
\begin{align}
\mathcal{H}_s &= \begin{bmatrix}
-J_z/2 & 0 & \sqrt{q-1}t & 0 & 0 & \dots \\ 
0 & -J_z/2 & t & 0 & 0 & \dots\\
\sqrt{q-1}t & t & (q-2)J_z/2 & \sqrt{q-1}t & 0 & \dots\\
0 & 0 & \sqrt{q-1}t & (q-2)J_z & 0 & \dots\\
0 & 0 & 0 & 0 & -J_z/2 & \dots \\
\vdots & \vdots & \vdots & \vdots & \vdots & \ddots
\end{bmatrix}, \\
\mathcal{H}_b &= \begin{bmatrix}
-J_z/2 & \sqrt{2(q-1)}t & 0 & 0 & \sqrt{q-1}t & \dots \\ 
\sqrt{2(q-1)}t & (q-2)J_z/2 & 0 & 0 & 0 & \dots\\
0 & 0 & -J_z/2 & \sqrt{2(q-2)}t & t & \dots\\
0 & 0 & \sqrt{2(q-2)}t & (q-2)J_z/2 & 0 & \dots\\
\sqrt{q-1}t & 0 & t & 0 & (q-2)J_z/2 & \dots \\
\vdots & \vdots & \vdots & \vdots & \vdots & \ddots
\end{bmatrix}.
\end{align}

\section{$p$ and $d$ waves} \label{app.p_d_waves}
In this section, we set up the $p$-wave and $d$-wave states for $q = 4$ nearest neighbors, define ordered bases for them, and compute the associated matrix representation of the Hamiltonian. The appropriate rotational eigenstates for the $p$-wave ($m = 1$) and $d$-wave ($m = 2$) states take on the forms
\begin{align}
\!\!\ket{\Psi_d^{m}(l_1,l_2)} &= \, [N^s_d(l_1,l_2)]^{-1/2}\!\!\!\sum_{\bj_{d+l_1+1}} {\rm e}^{-imj_1\pi/2} \!\!\! \sum_{\substack{\bi_{d+l_2+1}:\\\bi_d = \bj_d\\i_{d+1} \neq j_{d+1}}} \!\!\!\!\!\! \hat{h}^\dagger_{\bi_{d + l_2 + 1}} \hat{s}^\dagger_{\bj_{d}} \!\!\prod_{k_2=1}^{l_2}\!\!\hat{s}^\dagger_{\bi_{d + k_2}} \!\!\prod_{k_1=1}^{l_1}\!\!\hat{s}^\dagger_{\bj_{d + k_1}} \!\hat{h}^\dagger_{\bj_{d + l_1 + 1}}\!\ket{\AF},\! \nn\\
 \ket{\Psi^{m,c}_0(l_1,l_2)} &= [N_0(l_1,l_2)]^{-1/2}\!\!\sum_{\bj_{l_1+1}} {\rm e}^{-imj_1\pi / 2} \!\!\!\sum_{\substack{\bi_{l_2+1}:\\i_1 = j_1 + c}} \!\!\!\!\! \hat{h}^\dagger_{\bi_{l_2 + 1}} \hat{s}^\dagger_{0} \!\!\prod_{k_2=1}^{l_2}\hat{s}^\dagger_{\bi_{k_2}} \!\!\prod_{k_1=1}^{l_1}\hat{s}^\dagger_{\bj_{k_1}} \hat{h}^\dagger_{\bj_{l_1 + 1}}\ket{\AF}.
\label{app_eq.p_and_d_wave_states}
\end{align}
The upper line applies for $d \geq 1$, whereas the lower line are the more special $d = 0$ states in the case of $l_1 \neq l_2$. Here, $c = 1$ and $c = 2$ describe states on neighboring and opposite arms of the Bethe lattice as explained in the main text. We include $l_2 = -1$ in the $\ket{\Psi^{m,1}_0(l_1,l_2)}$ states for convinience. For $l_2 = l_1$ and $c = 2$, only the $p$-wave state
\begin{align}
\ket{\Psi^{1,2}_0(l_1,l_1)} &= [2N_0(l_1,l_1)]^{-1/2}\!\!\sum_{\bj_{l_1+1}} {\rm e}^{-ij_1\pi / 2} \!\!\!\sum_{\substack{\bi_{l_2+1}:\\i_1 = j_1 + c}} \!\!\!\!\! \hat{h}^\dagger_{\bi_{l_1 + 1}} \hat{s}^\dagger_{0} \!\!\prod_{k_2=1}^{l_1}\hat{s}^\dagger_{\bi_{k_2}} \!\!\prod_{k_1=1}^{l_1}\hat{s}^\dagger_{\bj_{k_1}} \hat{h}^\dagger_{\bj_{l_1 + 1}}\ket{\AF}.
\label{app_eq.p_wave_state_c_2_l2_equal_l1}
\end{align}
can be defined. The extra factor of $1/\sqrt{2}$ is to account for double counting. For the $d$-wave states this construction vanishes, because there is always two equivalent configurations with the opposite ordering of the holes. This also happens for the $p$ waves, but here there is an additional relative phase of ${\rm e}^{-i\pi} = -1$ that makes up for this. The normalization constant for the $d = 0$ states is
\begin{align}
N_0(l_1,l_2) = \sum_{\bj_{l_1+1}} \!\!\!\sum_{\substack{\bi_{l_2+1}:\\i_1 = j_1 + c}} 1 = q (q - 1)^{l_1}\left[\delta_{l_2,-1} + (1 - \delta_{l_2,-1}) (q - 1)^{l_2}\right],
\end{align}
where $q = 4$. We are now ready to write the ordered bases. These are
\begin{align}
{\bf p}\!: \big\{&\ket{\Psi^{1}_0(0,-1)}, \ket{\Psi^{1,1}_0(0,0)},.., \ket{\Psi^{1,1}_0(0,n)},\ket{\Psi^{1,2}_0(0,0)}, \nn\\
&\ket{\Psi^{1}_1(0,-1)}, \ket{\Psi^{1}_0(1,-1)}, \ket{\Psi^{1,1}_0(1,0)},.., \ket{\Psi^{1,1}_0(1,n)}, \ket{\Psi^{1,2}_0(1,0)},\ket{\Psi^{1,2}_0(1,1)}, .. \big\}. \nn \\
{\bf d}\!: \big\{&\ket{\Psi^{2}_0(0,-1)},\ket{\Psi^{2,1}_0(0,0)}, .., \ket{\Psi^{2,1}_0(0,n)}, \nn\\ 
&\ket{\Psi^{2}_1(0,-1)}, \ket{\Psi^{2}_0(1,-1)}, \ket{\Psi^{2,1}_0(1,0)}, .., \ket{\Psi^{2,1}_0(1,n)}, \ket{\Psi^{2,2}_0(1,0)}, .. \big\}.
\label{app_eq.ordered_bases_p_d}
\end{align}
As for the $s$ waves, we make overall blocks that have a constant value $d + l_1$ and running values of $l_1,l_2$ and subblocks with constant $l_1$ and running $l_2$. Note that the basis for the $p$ waves is larger, since $l_2 = l_1$ is allowed for the $\ket{\Psi^{1,2}_0(l_1,l_2)}$ states. The basis is again truncated by allowing a maximum value of $n = \max(d + l_1)$, setting a finite size of the system. The entry formulae now become
\begin{align}
{\bf p}\!: \; &j_p(d,l_1,l_2) = \frac{(d + l_1)(d + l_1 + 1)(d + l_1 + 2)}{6} + (d + l_1)(n + 2) + \frac{l_1(l_1 + 1)}{2} + l_2 + 1 + \delta_{c,1} + \delta_{c,2}(n+1), \nn \\
{\bf d}\!: \; &j_d(d,l_1,l_2) = \frac{(d + l_1)(d + l_1 + 1)(d + l_1 + 2)}{6} + (d + l_1)(n + 1) + \frac{l_1(l_1 + 1)}{2} + l_2 + 1 + \delta_{c,1} + \delta_{c,2}(n+2),
\label{app_eq.index_formula_p_d_wave}
\end{align}
which once again defines the entries of the matrix representations $\mathcal{H}_{p},\mathcal{H}_d$ of the Hamiltonian in each rotational subspace. The total size of the matrices are $N_r \times N_r$, with $N_p = j_p(0,n,n)$ and $N_d = j_d(0,n,n-1)$ in the two cases. We usually work with $n = 32$, corresponding to $N_p = 7667$ and $N_d = 7634$.  \\

As pointed out in the main text, any matrix elements away from $d = 0$ are equal to the ones for $s$ waves. We, therefore, focus solely on the $d = 0$ eigenstates in Eqs. \eqref{app_eq.p_and_d_wave_states} and \eqref{app_eq.p_wave_state_c_2_l2_equal_l1} here. There is only one instance, in which there is a coupling between $d = 0$ and $d\neq 0$ states. This is when the shallower hole follows the deeper hole as in Eq. \eqref{app_eq.d_and_l1_change_matrix_element_s_wave}, resulting in the same matrix element
\begin{equation}
\bra{\Psi^{m}_1(l_1-1,-1)} \hat{H}_t \ket{\Psi^{m}_0(l_1,-1)} = t \sqrt{\frac{N_1(l_1-1,-1)}{N_0(l_1,-1)}} = t. 
\end{equation}
The state $\ket{\Psi^{m}_0(l_1,-1)}$ can, furthermore, couple to $\ket{\Psi^{m,1}_0(l_1,0)}$ and $\ket{\Psi^{m,1}_0(0,l_1)}$ depending on, whether the shallower hole hops in front of the deeper hole ($i_1 > j_1$) or behind the deeper hole ($i_1 < j_1$). The associated matrix elements for $l_1 \geq 1$ are
\begin{align}
\bra{\Psi^{m,1}_0(l_1,0)} \hat{H}_t \ket{\Psi^{m}_0(l_1,-1)} &= t \sqrt{\frac{N_0(l_1,0)}{N_0(l_1,-1)}} = t,\nn \\ 
\bra{\Psi^{m,1}_0(0,l_1)} \hat{H}_t \ket{\Psi^{m}_0(l_1,-1)} &= -t \cdot \te^{-im\pi / 2} \sqrt{\frac{N_0(0,l_1)}{N_0(l_1,-1)}} = -t \cdot \te^{-im\pi / 2}.
\end{align}
The overall sign in the lower line arises due to exchange of the holes. The phase $\te^{-im\pi / 2}$ is there, because when the shallower hole hops behind the deeper hole, it corresponds to rotating the state $\ket{\Psi^{m,1}_0(0,l_1)}$ by $-\pi / 2$ radians. When $l_1 = 0$, these two events are indistinguishable and the total matrix element $\bra{\Psi^{m,1}_0(0,0)} \hat{H}_t \ket{\Psi^{m}_0(0,-1)} = t(1 - \te^{-im\pi / 2})$ describes the associated interference. Hopping deeper into the lattice from $l_2 \geq 0$, there is no interference, and we simply get
\begin{equation}
\bra{\Psi^{m,1}_1(l_1,l_2 + 1)} \hat{H}_t \ket{\Psi^{m,1}_0(l_1,l_2)} = t \sqrt{\frac{N_0(l_1,l_2 + 1)}{N_0(l_1,l_2)}} = \sqrt{3} t.
\end{equation}
using $q = 4$. If $l_2 = -1$, the shallower hole can also hop onto the opposite arm of the Bethe lattice. If $l_1 \geq 1$, this simply yields
\begin{equation}
\bra{\Psi^{m,2}_0(l_1,0)} \hat{H}_t \ket{\Psi^{m,1}_0(l_1,-1)} = t \sqrt{\frac{N_0(l_1,0)}{N_0(l_1,-1)}} = t. 
\end{equation}
If $l_1 = 0$, this hopping only happens for $p$ waves, where we get
\begin{equation}
\bra{\Psi^{1,2}_0(0,0)} \hat{H}_t \ket{\Psi^{1,1}_0(l_1,-1)} = t \sqrt{\frac{2N_0(l_1,0)}{N_0(l_1,-1)}} = \sqrt{2}t. 
\end{equation}
This describes a Bose-like enhanced hopping rate, where the fermionic antisymmetry is countered by the relative phase of the $p$-wave symmetry. This enhanced hopping happens for the $p$ waves, whenever the holes on opposite arms hop into the symmetrical configurations ($l_1 \geq 1$)
\begin{equation}
\bra{\Psi^{1,2}_0(l_1,l_1)} \hat{H}_t \ket{\Psi^{1,2}_0(l_1,l_1-1)} = t \sqrt{\frac{2N_0(l_1,l_1)}{N_0(l_1,l_1-1)}} = \sqrt{2}\cdot \sqrt{3}t.
\end{equation}
If $l_2 < l_1 - 1$, then we simply have $\bra{\Psi^{1,2}_0(l_1,l_2)} \hat{H}_t \ket{\Psi^{1,2}_0(l_1,l_2-1)} = \sqrt{3}t$. We now have all matrix elements in place. Below, we show the first $5\times 5$ entries
\begin{align}
\mathcal{H}_p &= \begin{bmatrix}
-J_z/2 & (1-i)t & 0 & 0 & 0 & \dots \\ 
(1+i)t & J_z & \sqrt{3}t & 0 & 0 & \dots\\
0 & \sqrt{3}t & 2J_z & 0 & 0 & \dots\\
0 & 0 & \sqrt{3}t & 3J_z & \sqrt{3}t & \dots\\
0 & 0 & 0 & \sqrt{3}t & 4J_z & \dots \\
\vdots & \vdots & \vdots & \vdots & \vdots & \ddots
\end{bmatrix}, \\
\mathcal{H}_d &= \begin{bmatrix}
-J_z/2 & 2t & 0 & 0 & 0 & \dots \\ 
2t & J_z & \sqrt{3}t & 0 & 0 & \dots\\
0 & \sqrt{3}t & 2J_z & 0 & 0 & \dots\\
0 & 0 & \sqrt{3}t & 3J_z & \sqrt{3}t & \dots\\
0 & 0 & 0 & \sqrt{3}t & 4J_z & \dots \\
\vdots & \vdots & \vdots & \vdots & \vdots & \ddots
\end{bmatrix}.
\end{align}
We note that that the first few entries of the Hamiltonian in the $p$- and $d$-wave subspaces are identical apart from $\mathcal{H}(1,2)$ describing the matrix element $\bra{\Psi^{m,1}_0(0,0)} \hat{H}_t \ket{\Psi^{m}_0(0,-1)}$. 

\section{Power-law behaviors in strongly correlated regime} \label{app.power_law_strong_correlations}
In this section, we briefly describe, how we extract the asymptotic power-law behaviors in the limit of strong correlations, $J_z \ll t$. \\

\begin{figure}[t!]
\begin{center}
\includegraphics[width=1.0\columnwidth]{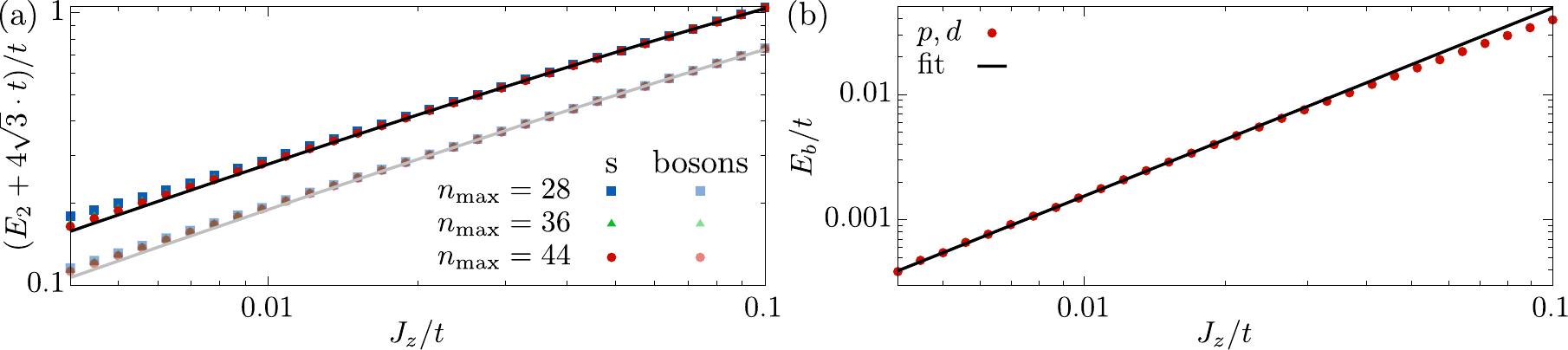}
\end{center}\vspace{-0.5cm}
\caption{(a) Exemplary data for the $s$-wave two-hole energy relative to the asymptotic value of $-4\sqrt{3}t$ for indicated values of the total depth of the lattice, $n_{\max}$. The upper points are for fermions, the lower points are for bosons. The asymptotically extracted fits are shown in black and grey lines, respectively. (b) $p$- and $d$-wave binding energies for $n_{\max} = 40$ along with the fit $E_b = 1.55t (J_z/t)^{3/2}$ [black line] showing excellent agreement.}
\label{fig.asymptotic_fits} 
\vspace{-0.25cm}
\end{figure}

For the $s$-wave fermionic and bosonic states we wish to extract the asymptotic form $E_b = 2E_1 - E_2 = c_{2/3} t (J_z / t)^{2/3} + c_1 J_z$. For a single hole, we previously found \cite{Nielsen2022_1}
\begin{align}
c_{2/3}^{(1)} &= a_0 \sqrt{q-1}\left(\frac{q-2}{2\sqrt{q-1}}\right)^{2/3} \overset{q = 4}{\simeq} 2.808, \nn \\
c_{1}^{(1)} &= \frac{1}{2}\left[1 + (q - 2) \left(d_0 + 2[\sqrt{q (q - 1)^{-1}} - 1] d_0 [1 - d_0] + [d_0 + 1/2]^2\right)\right] \overset{q = 4}{\simeq} -1.181.
\end{align}
Here, $-a_0 \simeq -2.33811$ is the first zero of the Airy function ${\rm Ai}(x)$. The upper line is an exact result, whereas the lower line is variationally determined in $d_0$, yielding $d_0 = 1/(2 - 3\sqrt{(q - 1)/q})$. The numerical values apply for $q = 4$. To determine the scaling of the binding energy, we make a fit of the form $E_2 + 4\sqrt{3}t = c_{2/3}^{(2)} t (J_z / t)^{2/3} + c_1^{(2)} J_z$ to the two-hole energy at low $J_z / t$ for a varying total depth of the lattice in the interval $n_{\max} \in [20,44]$. Performing a finite-size fit of $c_{j}^{(2)}$ then yields an approximate value for the infinite lattice case. The result of this fitting is shown in black and grey lines in Fig. \ref{fig.asymptotic_fits}(a) for the fermionic and bosonic case respectively. This is, furthermore, compared to exemplary data for indicated values of $n_{\max}$. We see excellent agreement between the data and the asymptotic fits. For the bosonic case, we get $c_{2/3}^{(2)} = 4.68$, $c_{1}^{(2)} = -2.75$. For the fermionic case, we get $c_{2/3}^{(2)} = 7.02$, $c_{1}^{(2)} = -4.77$. From here $c_{j} = 2c_j^{(1)} - c_j^{(2)}$, yielding
\begin{equation}
E_b^{(s)} = -1.40 t \left(\frac{J_z}{t}\right)^{2/3} + 2.40 J_z, \; E_b^{(b)} = 0.94 t \left(\frac{J_z}{t}\right)^{2/3} + 0.39 J_z,
\end{equation}
identical to Eq. \eqref{eq.binding_energy_strong_interactions}.\\

For the $p$ and $d$ waves, we expect that the binding energy is sublinear in $J_z$. This means that the two-hole binding energy must exactly cancel both the $(J_z/t)^{2/3}$ and linear $J_z$ term in $2E_1$. To test this, we fit a power-law to the binding energy $E_b = 2E_1 - E_2 = a\cdot t(J_z / t)^{b}$. In contrary to the $s$-wave states above, we find that this very quickly saturates as a function of the system size, yielding the same results for any $n_{\max} > 32$. Fitting to the $n_{\max} = 40$ data and guessing that the exponent is $3/2$, we find
\begin{equation}
E_b^{(p,d)} = 1.55 t \left(\frac{J_z}{t}\right)^{3/2},
\label{app_eq.binding_energy_p_d}
\end{equation}
Figure \ref{fig.asymptotic_fits}(b) shows excellent agreement with the sublinear behavior in Eq. \eqref{app_eq.binding_energy_p_d}.

%%%%%%%%%%%%%%%%%%%%%%%%%%%%%%%%%%%%%%%%%%%%%%%%%%%%%%%%%%%%%%%%%%% 
%%%%%%%%%%%%%%%%%%%%%%%%%%%%%%%%%%%%%%%%%%%%%%%%%%%%%%%%%%%%%%%%%%%

\newpage
\twocolumngrid

\bibliographystyle{apsrev4-2}
\bibliography{ref_magnetic_polaron}

\end{document}